# Spatially inhomogeneous inverse Faraday effect provides tunable nonthermal excitation of exchange dominated spin waves

Krichevsky D.M.[1,2*], Ozerov V.A.[1,2], Bel'kova A.V.[1,4], Sylgacheva D.A.[1], Kalish A.N.[1,4], Evstigneeva S.A.[1,5], Pakhomov A.S.[1,2], Mikhailova T.V.[3], Lyashko S.D.[3], Kudryashov A.L.[3], Semuk E.Yu.[3], Chernov A.I.[1,2], Berzhansky V.N.[3], Belotelov V.I.[1,3,4]

[1]Russian Quantum Center, 143025, Skolkovo, Moscow Region, Russia
[2]Moscow Institute of Physics and Technology (National Research University), 141700, Dolgoprudny, Russia
[3] V.I. Vernadsky Crimean Federal University, 295007, Simferopol, Russia
[4]Photonic and Quantum Technologies School, Faculty of Physics, Lomonosov Moscow State University, Moscow, Russia
[5]National University of Science and Technology, MISiS, Moscow 119991, Russia

*krichevskii.dm@phystech.edu

**We demonstrate optical nonthermal excitation of exchange dominated spin waves of different orders in a magnetophotonic crystal. The magnetophotonic structure consists of a thin magnetic film and a Bragg stack of nonmagnetic layers to provide a proper nonuniform interference pattern of the inverse Faraday effect induced by light in the magnetic layer. We found a phenomenon of the pronounced phase slippage of the inverse Faraday effect distribution when the pump wavelength is within the photonic band gap of the structure. It allows to tune the interference pattern by a slight variation of light wavelength which results in the modification of excitation efficiency of the different order spin waves. The approach can be applied for different magnetic dielectrics expanding their application horizons for spin-wave based devices.**

Efficient interaction between light and a magnet is vital for breakthrough technologies, such as magnetophotonics [1–5], all-optical energy-efficient magnetic recording [6] and spin-wave manipulation [7–9]. The latter one is an intrinsic part for spin-waves computation [10,11] in which high frequencies are of a great demand.

Precession of magnetic moments and propagation of these moments in the form of spin waves in magnetically ordered materials were first predicted almost a century ago by Bloch [12] and only a decade ago it was demonstrated that spatially shaped light can be used for the optical excitation of spin waves with the directional control [7]. The advances in ultrafast optical manipulation of magnetic order [6,13,14] provided new ways for the non-thermal excitation of propagating magnetostatic spin waves [7,15] in contrast to conventional microstrip antenna or coplanar waveguide [11,16,17], leading to the tunable energy-efficient spin wave manipulation [7,18] and even reconfigurable optomagnonic logic gates [19].

One way to optically and non-thermally excite and control spin waves is based on the inverse magnetooptical effects [20], in particular, the inverse Faraday effect (IFE). The effect occurs when a circularly polarized light pulse, passing through a magnetic medium, influences spins due to the stimulated Raman scattering, which is described in terms of an effective magnetic field induced by light. Experimentally, this method is implemented with the use of femtosecond laser pulses and it provides a substantial tunability of the excited spin waves properties in terms of their type [15], wavelength [21] and initial phase [22]. However, the most of the excited spin waves are described by magnetic dipolar interaction and are magnetostatic spin waves of $0^{th}$ order. In external magnetic fields of a moderate value their frequency is around few GHz.

Launching the higher frequency spin waves which are urgently awaited by magnonics is still scarcely achievable.

Wavelengths of high-frequency exchange-dominated spin waves are submicron. Therefore, a key point to efficiently excite them is to make a kind of nonuniformity of either internal magnetic field or an excitation stimulus inside a sample. The former approach was successfully applied for launching spin waves conventionally by microwaves in a periodically structured ferromagnet [23,24]. However, it is not possible to keep dealing with the microwave stimulus for uniform samples since microwave wavelengths far exceed their micron and especially nanometer thickness. One possibility is to use exchange torques [23,24], however it requires metallic parts of the structure and inevitably generates excess heating due to large amplitudes of the control electric currents.

Optical means are potentially very promising in this respect since wavelength of visible light in a magnet is comparable to the magnet thickness. A most straightforward way to achieve nonuniform impact of light on spins is due to the optical absorption which provides the exponential attenuation of the light intensity across the film thickness. It results to a nonuniform spin excitation profile due to modification of the magnetic anisotropy [25] or some other thermal effects [18]. Such method has been proven quite efficient and allowed to excite, for example, short exchange spin waves with supersonic velocities [18]. Nevertheless, such approach still requires light absorption and, therefore, has an inevitable side effect of undesirable heating and thermal losses.

The next step forward would be to avoid thermal mechanism of the optical excitation of exchange spin waves. Necessary nonuniform profile of the optical field in a magnet can be established not by the absorption of light but rather due to the optical interference or optical modes inside the sample. The latter has been implemented in an iron garnet film periodically perforated with nanotranches which allowed to excite nonthermally exchange spin waves of first two orders and switch between them by variation of angle of the linear polarized pump pulse [26]. In that experiment the nanotranches played a two-sided role: they provided an inhomogeneous distribution of both the optical energy and internal magnetic field across the sample. The latter complicates spin wave profile by evoking harmonics with lateral spatial distribution of spin dynamics.

Here we demonstrate nonthermal optical excitation of exchange-dominated standing spin waves (SSW) in a magnetophotonic crystal consisting of a smooth magnetic nanofilm covered with a dielectric nonmagnetic Bragg mirror. A periodic across film thickness spin excitation profile is achieved due to the interference of the incident laser pulse. Incident optical pump pulses are circularly polarized and therefore carry spin angular momentum and induce IFE effective magnetic field. Periodic distribution of this field with a period of 125-166 nm efficiently excites the exchange spin waves whose wavelength equals to the period of the optical pattern. By tuning excitation wavelength, we were able to adjust the IFE effective magnetic field distribution inside the magnetic layer and launch 3rd and 4th order exchange spin waves.

We performed the experiments on the optical excitation of SSW using 200 fs laser pulses whose wavelength is swept in a relatively short rage between $\lambda_{pm} = 610$ nm and $\lambda_{pm} = 685$ nm. Pump and probe pulses were generated by Newport Mai Tai Ti:sapphire laser (80.68 MHz repetition rate) combined with Spectra-Physics Inspire Auto 100 parametric oscillator. A delayed pump pulse was modulated using a photoelastic modulator (Hinds instruments PEM 100). The probe pulse (820 nm) was linearly polarized by a Glan-Taylor prism. Polarization changes of the probe pulse due to the Faraday effects were measured using auto-balanced optical receiver in the lock-in detection scheme. The sample was placed in the in-plane field of an electromagnet.

The considered magnetophotonic crystal consists of a 250 nm thick bismuth substituted iron garnet $(BiLu)_3Fe_5O_{12}$ (BIG) film grown on a $Gd_3Ga_5O_{12}$ paramagnetic substrate and covered by a dielectric $TiO_2/SiO_2$ (66/105 nm thick, respectively) Bragg mirror (Fig. 1a). It is placed in an external magnetic field which fully saturates the magnetic film in plane. Wavelength of the circularly polarized pump pulses is adjusted within the photonic band gap of the magnetophotonic crystal where the reflection coefficient is high (Fig. 1b). As a result, an interference pattern of the electromagnetic field of the laser pulse is formed. Since the pulse is circularly polarized it carries spin angular momentum and in accordance to the inverse Faraday effect induces the IFE effective magnetic field $\mathbf{H}_{IFE}$: $\mathbf{H}_{IFE} = -\frac{ig}{16\pi M_s}[\mathbf{E} \times \mathbf{E}^*]$ [20], where $g$ is gyration coefficient of the magnetic film, $M_s$ is its saturation magnetization, and $\mathbf{E}, \mathbf{E}^*$ are pump electric field vector and its complex conjugate. Due to the interference $\mathbf{H}_{IFE}$ becomes nonuniform across the film thickness and one should expect the optical pumping to excite standing spin waves whose wavelength and frequency are controlled by the pump wavelength.

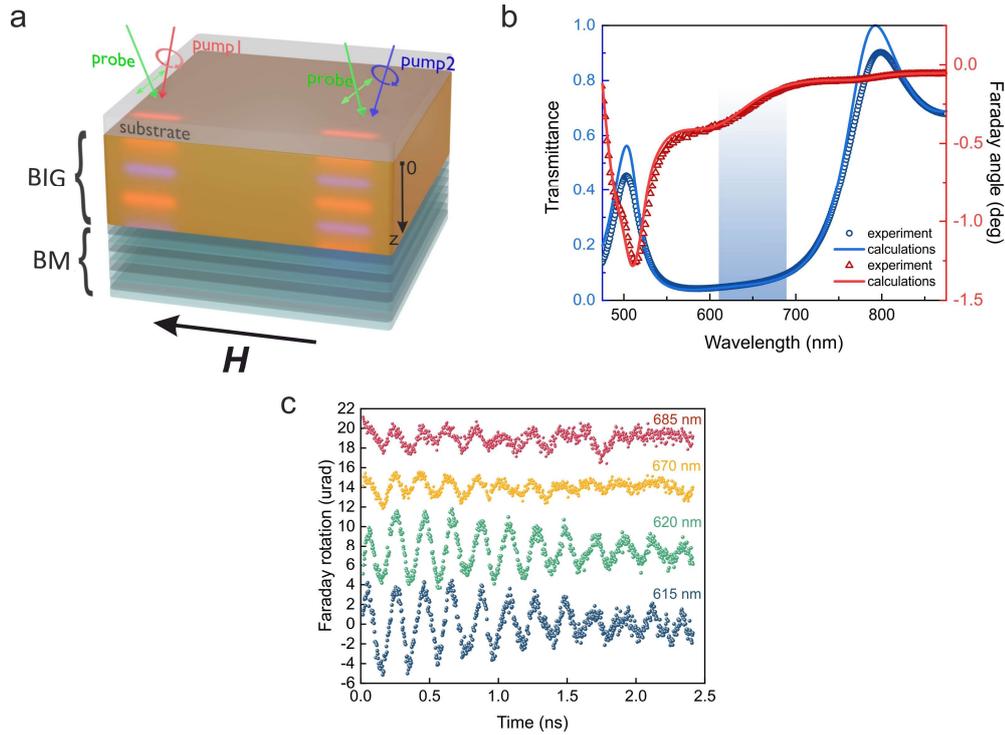

Fig. 1. The magnetophotonic crystal and optically excited spin dynamics. a – schematics of the magnetophotonic structure, consisting of a 250-nm thick bismuth iron-garnet layer sandwiched between TiO$_2$/SiO$_2$ Bragg mirror (BM) and GGG substrate. Bright maxima inside the BIG layer represent distribution of the IFE effective magnetic field induced by pumping at two different wavelengths: pump-1 (615 nm) and pump-2 (675 nm). b - transmission (blue) and Faraday rotation (red) spectra of the magnetophotonic crystal. The blue shaded area indicates range of the pump wavelengths. c – The spin precession launched by pump pulses of different wavelengths (615 nm, 625 nm, 670 nm and 685 nm) and observed by the Faraday rotation of the probe pulses at 970 Oe magnetic field.

The observed signals excited by different pumping wavelengths (Fig. 1c) have different spectral composition as follows from their Fourier spectra (Fig. 2a,b). For the pump wavelength of 610 nm the spectrum contains one pronounced peak at 4.8 GHz and a minor peak at 8.2 GHz. When the pump wavelength is increased a bit the primary peak remains the same while the second peak gets larger at $\lambda_{pm}$ =615 nm and then decreases. Therefore, pumping at $\lambda_{pm}$ =615 nm clearly provides two different spin modes: the low frequency one with large amplitude and the high frequency one with a smaller amplitude. Similar behavior takes place for the wavelength ranging from $\lambda_{pm} = 670$ nm to $\lambda_{pm} = 685$ nm: the main peak keeps at 4.8 GHz, while the subsidiary one increases for wavelength grows and maximizes at 685 nm. It should be noted that for this wavelength window the second peak appears at a frequency of 6.8 GHz which is smaller than for the secondary peak at the previous pumping wavelength range. Herein, the peak at 8.2 GHz disappears.

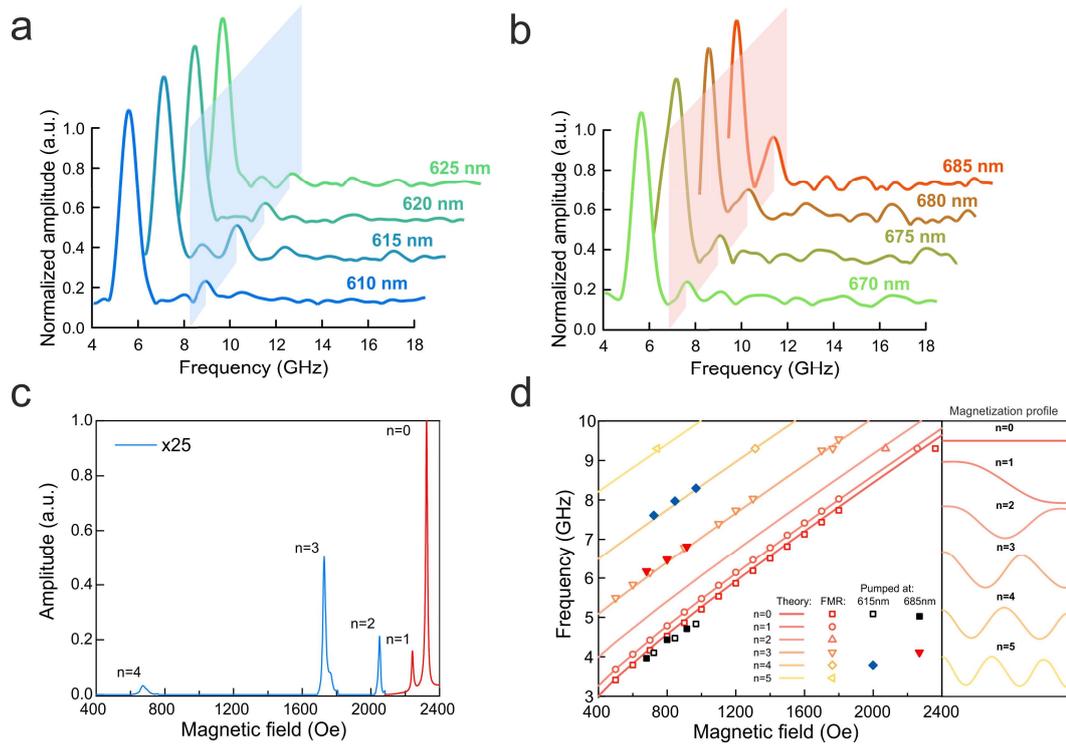

Fig. 2. Spectral properties of the optically excited spin oscillations and comparison to the microwave excitation (FMR measurements). a, b – FFT spectra of spin oscillations pumped at various wavelengths at 970 Oe external magnetic field. c – FMR spectra of the sample at three different magnetic fields and excitation frequency of 9.4 GHz. d – Dependance of the frequencies of different spin wave modes of the magnetic film excited by optical pumping (shaded symbols) and FMR (open symbols) on the external magnetic field. Solid curves represent corresponding calculated dependences (Supplementary, S1). Inset: schematic mode profiles of considered modes.

To shed a light on the observed oscillations origin and behavior we measured ferromagnetic resonance (FMR) in the sample. The investigation of magnetic properties of a magnetophotonic structure and the possibility of excitation of standing spin waves in the structure was carried out on a SPINSCAN EPR spectrometer at a fixed frequency of 9.4 GHz. In frequency-sweep VNA-FMR experiment the microwave absorption by the sample was measured at fixed external applied in-plane magnetic fields while driving the frequency. The frequency range of the sweep was from 2 to 8 GHz and the external magnetic field was varied from 900 to 2000 Oe.

Generally, since microwaves are uniformly distributed along the submicron thickness of the magnetic film it is very difficult to excite SSWs by them. However, in our FMR experiment the sample was placed in the microwave cavity which allowed us to detect SSWs of several orders though with very small amplitudes (Fig. 2c). Ratio of the amplitudes of the $0^{th}$-$1^{st}$-$2^{nd}$-$3^{rd}$-$4^{th}$ modes are 1:6:50:119:757, which confirms ultra-low excitation efficiency by microwave means. Figure 2d compares spectral positions of the detected FMR resonances (open symbols) and corresponding frequencies of the modes observed in the pump-probe experiment (shaded symbols, pump wavelength was set to 615 nm and 685 nm). Solid curves in Fig. 2d represent dependence of the calculated frequencies of SSW modes of different orders (solid curves) on magnetic field (Supplementary, section S1). It is seen that frequency of the low-frequency optically excited mode (grey and black shaded symbols) closely follows the frequency of the main FMR peak (open red symbols) and nicely corresponds to the theoretical curve for the uniform spin dynamics in the film ($n = 0$). Furthermore, the frequency of the high-frequency mode excited at 685 nm almost coincides with the fourth FMR peak and is in a good agreement with the calculated curve for the $3^{rd}$ order SSW. Similarly, the high-frequency mode excited at 615 nm is attributed to SSW of the $4^{th}$ order. Consequently, one can firmly conclude that, indeed, the modes observed in the pump-probe experiment are higher order exchanged spin waves. To study this phenomenon in detail let's consider the distribution of the IFE effective field inside the magnetic layer and its impact on the spin modes excitation.

Since the external magnetic field is saturating and is applied in-plane the n-th order SSW can be characterized by the out-of-plane dynamic magnetization component $M_{nz}(z,t)$. The SSW mode profile is given by $m_n(z)$ - the normalized amplitude of $M_{nz}(z,t)$. The excitation efficiency of the SSW modes is directly connected with an overlap between the IFE effective magnetic field distribution $H_{\text{IFE}}(z)$ and $n$-th order SSW mode profile [27–29] (Supplementary S3):

$$S_{\text{pm}} = \int_0^d H_{\text{IFE}}(z) m_n(z) dz \bigg/ \left| \int_0^d H_{\text{IFE}}(z) dz \right|, \quad (1)$$

where $d$ is the film thickness. Consequently, to maximize $S_{\text{pm}}$ it is crucial to adjust $H_{\text{IFE}}(z)$ to $m_n(z)$.

As light is incident close to the sample normal the optical interference forms a periodic pattern of the out-of-plain component of $H_{IFE}(z)$ across the film thickness (Supplementary S2):

$$H_{IFE}(z) \propto g[A_0(\lambda_{pm}) + A(\lambda_{pm}) \cos^2(k(\lambda_{pm})(z-d) + \varphi(\lambda_{pm}))], \quad (2)$$

where $A_0(\lambda_{pm})$, $A(\lambda_{pm})$ and $\varphi(\lambda_{pm})$ are parameters of the optical intensity distribution inside the BIG layer, $n_2$ is the refractive index of the magnetic layer, $k(\lambda_{pm}) = 2\pi n_2/\lambda_{pm}$, $d$ is thickness of the BIG layer, and $z$ is the coordinate normal to the film surface, $z = 0$ is at the interface between BIG layer and GGG substrate. The essential feature of the photonic crystal cover of the magnetic film is that for the photonic bandgaps $A_0(\lambda_{pm}) = 0$, so that

$$H_{IFE}(z) \propto gA(\lambda_{pm}) \cos^2\left(k(\lambda_{pm})(z-d) + \varphi(\lambda_{pm})\right), \quad (3)$$

where

$$\varphi(\lambda_{pm}) = \operatorname{atan}(-in_3/n_2), \quad (4)$$

$n_3$ is the complex effective refractive index of the Bragg mirror. Eq. (3) is illustrated by light blue and orange curves in Fig. 3a,b.

As it is seen from Eq. (3), both period and phase of $H_{IFE}(z)$ are sensitive to $\lambda_{pm}$. While the period of $H_{IFE}(z)$ is inversely proportional to $\lambda_{pm}$, the phase is almost linear in $\lambda_{pm}$ (Fig. 3c). This feature is unique for the photonic crystal cover as is illustrated in Fig. 3c. Fig. 3c shows the phase $\varphi(\lambda_{pm})$ deduced from the simulated optical field distributions for the case of the Bragg mirror (blue dots) and for the case when the Bragg mirror is replaced by air layer (red dots). The direct calculation of $\varphi(\lambda_{pm})$ by Eq. (4) is also shown (solid lines) (see Supplementary S2 for details). This effect of strong phase slippage $\varphi(\lambda_{pm})$ of the IFE appears in the magnetophotonic crystal for the wavelengths inside the photonic band gap and is crucial for SSW control as will be discussed below.

As for $m_n(z)$ distribution, the 0$^{\text{th}}$ order SSW mode is quasi uniform across the magnetic film ($m_0(z) \approx$ const), whereas higher order modes have profiles close to sine or cosine functions (blue and red curves in Fig. 3a,b). Having uniform distribution, the 0$^{\text{th}}$-mode is excited by both uniform and nonuniform IFE fields since the pumping efficiency $S_{\text{pm}}$ remains relatively large for these cases. On the contrary, as it follows from Eq.(1), the higher spin modes require periodic $H_{\text{IFE}}(z)$ with the period comparable to their wavelength. The excitation efficiency (namely, modulus of the amplitude) of the 3$^{\text{rd}}$ and 4$^{\text{th}}$ order SSWs as a function of pump wavelength is summarized in Fig. 3d. The 4$^{\text{th}}$ order SSWs is excited most efficiently at 615-620 nm, where its profile almost coincides to the profile of $H_{\text{IFE}}(z)$ (blue curve and symbols in Fig. 3d). Increase of the pump wavelength detunes $\mathbf{H}_{\text{IFE}}(z)$ profile from the 4$^{\text{th}}$-SSW one and amplitude of 4$^{\text{th}}$-SSW drops down almost to zero at $\lambda_{pm} = 695$ nm. At the same time, $H_{\text{IFE}}(z)$ profile gets closer to the 3$^{\text{rd}}$-SSW one and amplitude of 3$^{\text{rd}}$-SSW grows (red curve and symbols in Fig. 3d). It reaches parity with 4$^{\text{th}}$-SSW at $\lambda_{pm} = 660$ nm and grows further up to $\lambda_{pm} = 685$ nm where it maximizes. Therefore, changing pump wavelength, one could switch between different orders of exchange spin waves or excite their superposition.

Tuning of $H_{\text{IFE}}(z)$ between optimum for excitation of 4th-SSW and 3rd-SSW is possible due to dependence of its spatial period and, most importantly, phase on the pump wavelength: $k(\lambda_{pm})$ and $\varphi(\lambda_{pm})$ (see Eq.(2)). It is important to note the role of the photonic crystal. Indeed, if the photonic crystal was replaced by a dielectric layer, then $\varphi(\lambda_{pm})$ would be constant and to tune between 4th-SSW and 3rd-SSW one would need to change $k(\lambda_{pm})$ substantially, by $\Delta k = \pi/2d$, which would require $\Delta\lambda_{pm} = dn_2/3 = 207$ nm. Here the presence of the photonic crystal provides $\varphi(\lambda_{pm})$ and one has to increase the pump wavelength by only $\Delta\lambda_{pm} = dn_2/3 = 75$ nm to get the same level of tunability.

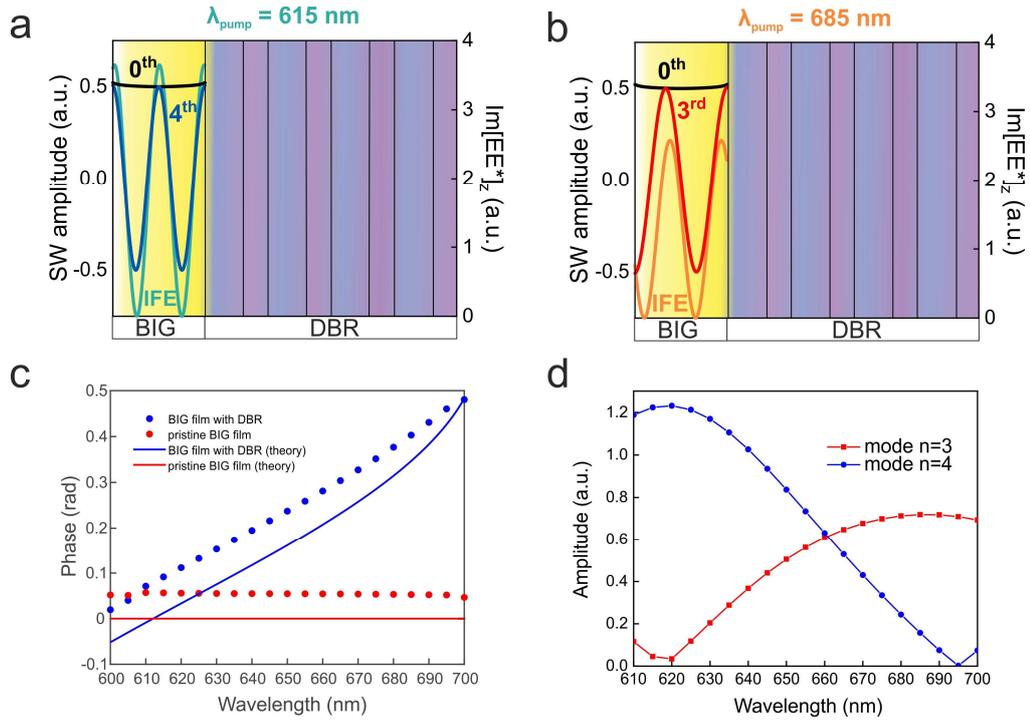

Fig. 3. Profiles of the SSWs. a, b – numerically calculated IFE field distribution inside the structure at 615 and 685 nm and profiles of the SSW modes of 0th, 3rd and 4th order. c - the phase of the IFE field inside the BIG layer as a function of pump wavelength for pristine (red color) and photonic crystal (blue color) covered films. Dots show the phase deduced from the simulated field distribution, lines give the calculations by Eq. (4) (see Supplement, S2 for details). d – numerically calculated dependence of the third and fourth standing spin wave modes amplitude as a function of wavelength. Magnetic field is 980 Oe.

Let's now discuss relative values of the SSW mode amplitudes observed in the experiment. It is important to note that the magneto-optically probed SSW amplitudes might be significantly different from the SSW amplitudes in reality. The Faraday effect, which is widely employed for probing in the ultrafast experiments [20,30], is sensitive to the dynamic component of magnetization which is along wavevector of the probe. If the probe pulse shines the sample at normal or close to normal incidence then the Faraday effect is proportional to the out-of-plane component of magnetization. In case of an SSW mode $m_n(z)$ and the probe pulse intensity $W_{\text{pr}}(z)$ are nonuniformly distributed across the film thickness. Consequently, the Faraday angle is proportional to the maximal value of the out-of-plane magnetization of $n^{th}$ mode, $M_{nz}$, and to the probing efficiency $S_{\text{pr}} = \int_0^h W_{\text{pr}}(z) m_n(z) dz / \left|\int_0^h W_{\text{pr}}(z) dz\right|$. For the case of 0th order mode typically observed in thin films, magnetization distribution remains quasi-homogeneous ($m_0(z) \approx$ const). Similarly, distribution of the probe in a magnetic film is also close to the uniform one ($W_{\text{pr}}(z) \approx$ const), which makes detection efficiency of this mode quite large ($S_{\text{pr}} \approx 1$). In the case of the magnetophotonic crystal pump and probe become non-uniform across the film and, as we discussed above, the non-uniform pumping launches higher order modes (see Eq.(2)) whose amplitude is also thickness dependent ($m_n(z)$). As a result, the Faraday signal of the probe beam is determined by not only a value of $M_{nz}$ but also by relative distribution of $W_{\text{pr}}(z)$ and $m_n(z)$. In some cases, the observed Faraday effect could be quite small even for large $M_{nz}$. Following these ideas, we numerically calculated (Supplementary S1) real amplitudes of the SSW modes (transparent bars in Fig. 4) and the amplitudes of the SSW modes detected by the Faraday rotation of 820 nm probe pulses (shaded bars in Fig. 4).

A 615 nm pump pulse excites $0^{\text{th}}$ and $4^{\text{th}}$ order SSW modes with similar amplitudes. Moreover, the amplitude of the $4^{\text{th}}$-mode is even a bit larger (see transparent bars in Fig. 4a). However, the observed relative amplitudes of the $0^{\text{th}}$ and $4^{\text{th}}$ modes are quite different, their ratio is around 20, since for the $0^{\text{th}}$-mode the detection integral $S_{\text{pr}}$ is much larger (see shaded bars in Fig. 4a). A similar situation is noticed for the 685 nm pump which excites mostly $0^{\text{th}}$ and $3^{\text{rd}}$ SSW modes: the observed ratio of the mode amplitudes is 1:5 (shaded bars in Fig. 4b) while in reality their amplitudes are almost the same (transparent bars in Fig. 4b). The differences in mode amplitudes between experimental and numerically calculated values are primarily due to incident angle variation for both pump and probe beams, as well as structure inhomogeneities caused by the fabrication process.

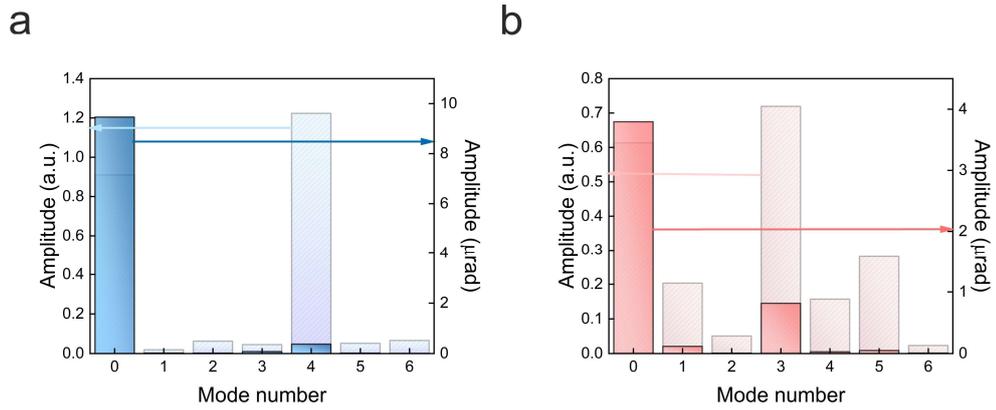

Fig. 4. Amplitudes of the standing spin wave modes pumped at 615 (a) and 685 nm (b): real (transparent bars) and as observed by the probe at 820 nm (shaded bars). Magnetic field is 980 Oe.

Our experiments demonstrate spectrally adjustable excitation of high-order SSW caused by inhomogeneous effective magnetic field formed inside the magnetic layer of the magnetophotonic structure by femtosecond laser pulses. The magnetophotonic crystal provides necessary distribution of the inverse Faraday effect magnetic field and also gives a high level of tunability since spatial phase of the inverse Faraday effect inside its magnetic layer becomes strongly dependent on the pump wavelength. This effect appears in the photonic band gap of the structure. As a result, $3^{\text{rd}}$ and $4^{\text{th}}$ order spin modes are shown to be induced with an efficiency close to the $0^{\text{th}}$ mode efficiency which is not reachable by microwave means. To tune between these modes variation of the pump wavelength by only 75nm is enough. However, visibility of the spin modes in the experiment is several times lower due to not optimal distribution of the probe pulse across the film thickness. To increase the probing efficiency the probe wavelength should be chosen to get distribution of the probe beam close the pump one. Optical excitation of high-order exchange spin modes is of prime importance for data processing applications involving optospintronic and optomagnonic devices with higher-frequencies demands.

**Acknowledgments**

This work was financially supported by the Ministry of Science and Higher Education of the Russian Federation, Megagrant project N 075-15-2022-1108.

**References**

1. B. Wang, K. Rong, E. Maguid, V. Kleiner, and E. Hasman, "Probing nanoscale fluctuation of ferromagnetic meta-atoms with a stochastic photonic spin Hall effect," Nat Nanotechnol **15**, 450–456 (2020).

2. V. V. Temnov, G. Armelles, U. Woggon, D. Guzatov, A. Cebollada, A. Garcia-Martin, J.-M. Garcia-Martin, T. Thomay, A. Leitenstorfer, and R. Bratschitsch, "Active magneto-plasmonics in hybrid metal–ferromagnet structures," Nat Photonics **4**, 107–111 (2010).

3. J. Kuttruff, A. Gabbani, G. Petrucci, Y. Zhao, M. Iarossi, E. Pedrueza-Villalmanzo, A. Dmitriev, A. Parracino, G. Strangi, F. De Angelis, D. Brida, F. Pineider, and N. Maccaferri, "Magneto-Optical Activity in Nonmagnetic Hyperbolic Nanoparticles," Phys Rev Lett **127**, 217402 (2021).

4. M. J. Steel, M. Levy, and R. M. Osgood, "High transmission enhanced Faraday rotation in one-dimensional photonic crystals with defects," IEEE Photonics Technology Letters **12**, 1171–1173 (2000).

5. M. Inoue, K. Arai, T. Fujii, and M. Abe, "Magneto-optical properties of one-dimensional photonic crystals composed of magnetic and dielectric layers," J Appl Phys **83**, 6768–6770 (1998).

6. A. Stupakiewicz, K. Szerenos, D. Afanasiev, A. Kirilyuk, and A. V. Kimel, "Ultrafast nonthermal photo-magnetic recording in a transparent medium," Nature **542**, 71–74 (2017).

7. T. Satoh, Y. Terui, R. Moriya, B. A. Ivanov, K. Ando, E. Saitoh, T. Shimura, and K. Kuroda, "Directional control of spin-wave emission by spatially shaped light," Nat Photonics **6**, 662–666 (2012).

8. M. van Kampen, C. Jozsa, J. T. Kohlhepp, P. LeClair, L. Lagae, W. J. M. de Jonge, and B. Koopmans, "All-Optical Probe of Coherent Spin Waves," Phys Rev Lett **88**, 227201 (2002).

9. P. I. Gerevenkov, Ia. A. Filatov, A. M. Kalashnikova, and N. E. Khokhlov, "Unidirectional Propagation of Spin Waves Excited by Femtosecond Laser Pulses in a Planar Waveguide," Phys Rev Appl **19**, 024062 (2023).

10. D. Sander, S. O. Valenzuela, D. Makarov, C. H. Marrows, E. E. Fullerton, P. Fischer, J. McCord, P. Vavassori, S. Mangin, P. Pirro, B. Hillebrands, A. D. Kent, T. Jungwirth, O. Gutfleisch, C. G. Kim, and A. Berger, "The 2017 Magnetism Roadmap," J Phys D Appl Phys **50**, 363001 (2017).

11. B. Lenk, H. Ulrichs, F. Garbs, and M. Münzenberg, "The building blocks of magnonics," Phys Rep **507**, 107–136 (2011).

12. F. Bloch, "Zur Theorie des Ferromagnetismus," Zeitschrift für Physik **61**, 206–219 (1930).

13. A. V. Kimel, A. Kirilyuk, P. A. Usachev, R. V. Pisarev, A. M. Balbashov, and Th. Rasing, "Ultrafast non-thermal control of magnetization by instantaneous photomagnetic pulses," Nature **435**, 655–657 (2005).

14. C.-H. Lambert, S. Mangin, B. S. D. Ch. S. Varaprasad, Y. K. Takahashi, M. Hehn, M. Cinchetti, G. Malinowski, K. Hono, Y. Fainman, M. Aeschlimann, and E. E. Fullerton, "All-optical control of ferromagnetic thin films and nanostructures," Science (1979) **345**, 1337–1340 (2014).

15. A. I. Chernov, M. A. Kozhaev, I. V. Savochkin, D. V. Dodonov, P. M. Vetoshko, A. K. Zvezdin, and V. I. Belotelov, "Optical excitation of spin waves in epitaxial iron garnet films: MSSW vs BVMSW," Opt Lett **42**, 279 (2017).

16. C. Liu, J. Chen, T. Liu, F. Heimbach, H. Yu, Y. Xiao, J. Hu, M. Liu, H. Chang, T. Stueckler, S. Tu, Y. Zhang, Y. Zhang, P. Gao, Z. Liao, D. Yu, K. Xia, N. Lei, W. Zhao, and M. Wu, "Long-distance propagation of short-wavelength spin waves," Nat Commun **9**, 738 (2018).

17. S. Neusser and D. Grundler, "Magnonics: Spin Waves on the Nanoscale," Advanced Materials **21**, 2927–2932 (2009).

18. J. R. Hortensius, D. Afanasiev, M. Matthiesen, R. Leenders, R. Citro, A. V. Kimel, R. V. Mikhaylovskiy, B. A. Ivanov, and A. D. Caviglia, "Coherent spin-wave transport in an antiferromagnet," Nat Phys **17**, 1001–1006 (2021).

19. A. A. Kolosvetov, M. A. Kozhaev, I. V. Savochkin, V. I. Belotelov, and A. I. Chernov, "Concept of the Optomagnonic Logic Operation," Phys Rev Appl **18**, 054038 (2022).

20. A. Kirilyuk, A. V. Kimel, and T. Rasing, "Ultrafast optical manipulation of magnetic order," Rev Mod Phys **82**, 2731–2784 (2010).

21. M. Jäckl, V. I. Belotelov, I. A. Akimov, I. V. Savochkin, D. R. Yakovlev, A. K. Zvezdin, and M. Bayer, "Magnon Accumulation by Clocked Laser Excitation as Source of Long-Range Spin Waves in Transparent Magnetic Films," Phys Rev X **7**, 021009 (2017).

22. A. I. Chernov, M. A. Kozhaev, A. Khramova, A. N. Shaposhnikov, A. R. Prokopov, V. N. Berzhansky, A. K. Zvezdin, and V. I. Belotelov, "Control of the phase of the magnetization precession excited by circularly polarized femtosecond-laser pulses," Photonics Res **6**, 1079 (2018).

23. A. Navabi, C. Chen, A. Barra, M. Yazdani, G. Yu, M. Montazeri, M. Aldosary, J. Li, K. Wong, Q. Hu, J. Shi, G. P. Carman, A. E. Sepulveda, P. Khalili Amiri, and K. L. Wang, "Efficient Excitation of High-Frequency Exchange-Dominated Spin Waves in Periodic Ferromagnetic Structures," Phys Rev Appl **7**, 034027 (2017).

24. H. Qin, S. J. Hämäläinen, and S. van Dijken, "Exchange-torque-induced excitation of perpendicular standing spin waves in nanometer-thick YIG films," Sci Rep **8**, 5755 (2018).

25. M. Deb, E. Popova, M. Hehn, N. Keller, S. Petit-Watelot, M. Bargheer, S. Mangin, and G. Malinowski, "Femtosecond Laser-Excitation-Driven High Frequency Standing Spin Waves in Nanoscale Dielectric Thin Films of Iron Garnets," Phys Rev Lett **123**, 027202 (2019).

26. A. I. Chernov, M. A. Kozhaev, D. O. Ignatyeva, E. N. Beginin, A. V. Sadovnikov, A. A. Voronov, D. Karki, M. Levy, and V. I. Belotelov, "All-Dielectric Nanophotonics Enables Tunable Excitation of the Exchange Spin Waves," Nano Lett **20**, 5259–5266 (2020).

27. V. A. Ozerov, D. A. Sylgacheva, M. A. Kozhaev, T. Mikhailova, V. N. Berzhansky, M. Hamidi, A. K. Zvezdin, and V. I. Belotelov, "One-dimensional optomagnonic microcavities for selective excitation of perpendicular standing spin waves," J Magn Magn Mater **543**, 168167 (2022).

28. D. M. Krichevsky, D. O. Ignatyeva, V. A. Ozerov, and V. I. Belotelov, "Selective and Tunable Excitation of Standing Spin Waves in a Magnetic Dielectric Film by Optical Guided Modes," Phys Rev Appl **15**, 034085 (2021).

29. A. G. Gurevich and G. A. Melkov, *Magnetization Oscillations and Waves* (CRC Press, 2020).

30. P. Němec, M. Fiebig, T. Kampfrath, and A. V. Kimel, "Antiferromagnetic opto-spintronics," Nat Phys **14**, 229–241 (2018).

# Supplemental document

# Spatially inhomogeneous inverse Faraday effect provides tunable nonthermal excitation of exchange dominated spin waves

Krichevsky D.M.[1,2], Bel'kova A.V.[1,4], Ozerov V.A.[1,2,3], Sylgacheva D.A.[1], Kalish A.N.[1,4], Evstigneeva S.A.[1,5], Pakhomov A.S.[1,2], Mikhailova T.V.[3], Lyashko S.D.[3], Kudryashov A.L.[3], Semuk E.Yu.[3], Chernov A.I.[1,2,3], Berzhansky V.N.[3], Belotelov V.I.[1,3,4]

[1] Russian Quantum Center, 143025, Skolkovo, Moscow Region, Russia
[2] Moscow Institute of Physics and Technology (National Research University), 141700, Dolgoprudny, Russia
[3] V.I. Vernadsky Crimean Federal University, 295007, Simferopol, Russia
[4] Photonic and Quantum Technologies School, Faculty of Physics, Lomonosov Moscow State University, Moscow, Russia
[5] National University of Science and Technology, MISiS, Moscow 119991, Russia

**S1. Landau-Lifshitz-Gilbert equation and SSW.**

Magnetization dynamics launched by ultrashort laser pulses can be investigated on the basis of the Landau-Lifshitz-Gilbert equation:

$$\frac{d\mathbf{M}}{dt} = -\gamma[\mathbf{M} \times \mathbf{H}_{eff}] + \frac{\alpha}{|\mathbf{M}|}\left[\mathbf{M} \times \frac{d\mathbf{M}}{dt}\right]. \quad (S1.1)$$

Here the effective magnetic field acting on magnetization is $\mathbf{H}_{eff} = \mathbf{H} + 4\pi\mathbf{M} - \frac{2K_U}{M}\mathbf{e}_z + A\Delta\mathbf{M} + \mathbf{H}_{\text{IFE}}$, where $K_U$ is uniaxial anisotropy constant, $A$ is exchange constant, $\gamma$ is gyromagnetic ratio and $\alpha$ is Gilbert damping constant considered to be small ($\alpha \sim 10^{-3}$). The direction of the external magnetic field $\mathbf{H} = \mathbf{e}_x H$ corresponds to the x-axis (which lies in a sample plane), while the IFE-field $\mathbf{H}_{\text{IFE}} = \mathbf{e}_z f(t)h(z)$ is oriented along z-axis (which is perpendicular to the sample plane). We can define the spherical coordinate system in terms of $\theta, \varphi$ angles, where $\theta$ is a deflection angle between the magnetization $\mathbf{M}$ and its in-plane projection $\mathbf{M}_{xy}$, whereas $\varphi$ is a deflection angle between the in-plane projection $\mathbf{M}_{xy}$ and the external field $\mathbf{H}$. In this coordinate system we have $M_y = |\mathbf{M}|\cos\theta\sin\varphi$; $M_z = |\mathbf{M}|\sin\theta$ and Eq. (S1.1) is rewritten as:

$$\begin{cases} \dot{\theta} = \alpha\dot{\varphi} + \gamma H\varphi - \gamma AM\varphi'', \\ \dot{\varphi} = -\alpha\dot{\theta} - \gamma\widetilde{H}\theta + \gamma AM\theta'' + \gamma f(t)h(z). \end{cases} \quad (S1.2)$$

Here we assumed that the precession angle is small ($\theta \ll 1$, $\varphi \ll 1$), and the derivatives are denoted by $\theta'' = \frac{\partial^2 \theta}{\partial z^2}$, $\varphi'' = \frac{\partial^2 \varphi}{\partial z^2}$, $\dot{\theta} = \frac{\partial \theta}{\partial t}$, $\dot{\varphi} = \frac{\partial \varphi}{\partial t}$. For convenience, we introduced the notation $\widetilde{H} = H + 4\pi M - \frac{2K_U}{M}$, where $H = |\mathbf{H}|$, $M = |\mathbf{M}|$.

The term $\gamma f(t)h(z)$ in Eq. (S1.2) (here $h(z)$ corresponds to the average value of the magnetic field of the inverse Faraday effect during the pulse propagation) is non-zero only during the small time of laser pulse propagation $\Delta t$, meaning that it is responsible only for the establishing of the initial conditions for $\theta, \varphi$. If we integrate Eq. (S1.2) by $t$ from 0 to $\Delta t$ using $\int_0^{\Delta t} f(t)\, dt = \Delta t$, we will find the initial conditions for $\theta, \varphi$ after the instant stimulus of a laser pump:

$$\begin{cases} \varphi(z, t = \Delta t) = \gamma \Delta t \cdot h(z), \\ \theta(z, t = \Delta t) \cong 0. \end{cases} \quad (S1.3)$$

Besides, boundary conditions should also be taken into consideration. As shown in [1], for the given configuration the boundary conditions take the form:

$$\begin{cases} \theta' + \xi\theta = 0; & z = 0, \\ \theta' - \xi\theta = 0; & z = d, \\ \varphi' = 0; & z = 0, \\ \varphi' = 0; & z = d. \end{cases} \quad (S1.4)$$

Here $\xi$ is a pinning parameter, originating from the surface anisotropy, $d$ is the thickness of the magnetic film. Parameter $\xi$ may be expressed in terms of the surface anisotropy parameter $K_s$ as follows: $\xi = \frac{2K_s}{AM^2}$. For the further calculations we assumed $\xi d = 0.5$ (the general case of partially pinned spins).

Eq. (S1.2) together with Eqs. (S1.3) and (S1.4) fully formulates the Cauchy differential equation problem. The solution of Eq. (S1.2) has the form of decaying harmonic oscillations $\theta(z,t), \varphi(z,t) \sim e^{i(kz-\omega t)-\lambda t}$. Here, assuming that $\lambda$ (which is proportional to $\alpha$) is small, the frequency $\omega$ is expressed through the wavevector $k$ as follows: $\omega^2 = \gamma^2(H + AMk^2)(\widetilde{H} + AMk^2)$. We should note that two values of wavevector $k$ ($k_+$ and $k_-$) correspond to the given value of the frequency $\omega$. The first one ($k_+$) can be either real ($k_+ = k$) or imaginary ($k_+ = i\chi_+$), depending on $\omega$, and the second one ($k_-$) is always imaginary ($k_- = i\chi$). Solutions with imaginary $k_\pm$ are responsible for hyperbolic (surface) terms of the modes, and can't be neglected. The more detailed explanation was given in Ref. [2].

Since the excited PSSW oscillations can be detected through the Faraday effect, which is sensitive to the normal component of the magnetization, we will describe the PSSWs by the $\theta$ angle. There are two types of the PSSW modes $\theta_n(z,t) = \theta_n(t) \cdot \theta_n(z)$:

$$\theta_n(z,t) = e^{-\lambda_n t} \sin \omega_n t \cdot \begin{cases} \cos k_n z' + b_n B_n \cosh \chi_n z', & n = 2,4,6,\ldots \\ \sin k_n z' + b_n B_n \sinh \chi_n z', & n = (1),3,5,\ldots \end{cases}, \quad (S1.5a)$$

$$\theta_n(z,t) = e^{-\lambda_n t} \sin \omega_n t \cdot \begin{cases} \cosh \chi_{+,n} z' + \tilde{b}_n \tilde{B}_n \cosh \chi_n z', & n = 0 \\ \sinh \chi_{+,n} z' + \tilde{b}_n \tilde{B}_n \sinh \chi_n z', & n = (1) \end{cases}. \quad (S1.5b)$$

Here, for simplicity, we introduced the notation: $z' = z - \frac{d}{2}$.

However, the expressions for modes $\varphi_n(z,t) = \varphi_n(t) \cdot \varphi_n(z)$ will also play the role for the future calculations:

$$\varphi_n(z,t) = e^{-\lambda_n t} \cos \omega_n t \cdot \sqrt{-b_n} \begin{cases} \cos k_n z' + B_n \cosh \chi_n z', & n = 2,4,6,\ldots \\ \sin k_n z' + B_n \sinh \chi_n z', & n = (1),3,5,\ldots \end{cases}, \quad (S1.6a)$$

$$\varphi_n(z,t) = e^{-\lambda_n t} \cos \omega_n t \cdot \sqrt{-b_n} \begin{cases} \cosh \chi_{+,n} z' + \tilde{B}_n \cosh \chi_n z', & n = 0 \\ \sinh \chi_{+,n} z' + \tilde{B}_n \sinh \chi_n z', & n = (1) \end{cases}. \quad (S1.6b)$$

Here, the modes symmetric with respect to the film center correspond to even $n$, the antisymmetric modes – to the odd $n$. Note, that the modes with $n = 1$ may have a form of Eqs. (S1.5a)-(S1.6a) or Eqs. (S1.5b)-(S1.6b) depending on the product $\xi d$. For small product $\xi d$ the $n = 1$ mode will take a form of Eqs. (S1.5a)-(S1.6a).

It should also be noted that for small values of $\xi d$ hyperbolic terms in Eqs. (S1.5a)-(S1.6a) are relatively small and these modes can be considered as «quasi-harmonic» with wavevector $k_n \approx \frac{\pi n}{d} - \frac{\xi}{\pi n}$. As for the modes described by Eqs. (S1.5b)-(S1.6b), we may call them «hyperbolic» modes, as they are expressed through the sum of hyperbolic functions. The expressions for all the coefficients and wavevectors presented in Eqs.(S1.5)-(S1.6) are described in more details in Ref. [20].

A spatially non-uniform instant stimulus will excite a set of eigenmodes with different amplitudes $A_n$:

$$\theta(z,t) = \sum_{n=0}^{\infty} A_n \cdot \theta_n(z,t). \quad (S1.7)$$

These amplitudes $A_n$ depend on the distribution of the IFE-field $h(z)$, which determines the initial conditions, and may be found using the following expression:

$$A_n = \frac{\gamma \Delta t}{\left(\int_0^d \varphi_n(z) \cdot \theta_n(z) \, dz\right)} \int_0^d h(z) \cdot \theta_n(z) \, dz. \quad (S1.8)$$

The detailed derivation of Eq. (S1.8) is given in S3.

## S2. Optical field distribution inside the BIG layer.

Let's consider the case of normal incidence of light on a BIG layer. In this case, the effective field $\mathbf{H}_{IFE} = -\frac{ig}{16\pi M_s}[\mathbf{E} \times \mathbf{E}^*]$ is obviously directly proportional to the optical field intensity: $H_{IFE}(z) \propto g|E(z)|^2$.

If the BIG layer is surrounded by two media, the optical field distribution has the form:

$$E(z) = A\{n_2 \cos(n_2 k_0 (z-d)) + i n_3 \sin(n_2 k_0 (z-d))\}, \quad (S2.1)$$

where

$$A = \frac{2n_1}{n_2(n_1 + n_3)\cos(n_2 k_0 d) - i(n_2^2 + n_1 n_3)\sin(n_2 k_0 d)}. \quad (S2.2)$$

Here $n_1$ is the refractive index of the incoming medium, $n_2$ and $n_3$ are refractive indices of the BIG film and the backward medium, respectively, $i$ is the imaginary unit, $k_0$ is the vacuum wavelength, $z$ is the coordinate normal to the film surface, $z = 0$ is at the interface between the BIG layer and the incoming medium. It can be easily shown that for the transparent media the intensity distribution has the form

$$|E(z)|^2 = A_0 + A_1 \cos^2(n_2 k_0 (z-d) + \varphi), \quad (S2.3)$$

which corresponds to Eq. (2).

Eq. (S2.1) remains valid for the case when the backward medium is replaced by the semi-infinite photonic crystal. In this case $n_3$ is the complex effective refractive index of the photonic crystal defined as the relation between magnetic and electric fields of the Bloch wave at the interface:

$$n_3 = \frac{H(z=0)}{E(z=0)}, \quad (S2.4)$$

From the explicit equations for the Bloch wave fields one can obtain:

$$n_3 = i n_4 \frac{2n_5 \exp(iK(a+b)) - (n_4 + n_5)\cos(k_0 n_4 a + k_0 n_5 b) + (n_4 - n_5)\cos(k_0 n_4 a - k_0 n_5 b)}{(n_4 - n_5)\sin(k_0 n_4 a - k_0 n_5 b) - (n_4 + n_5)\sin(k_0 n_4 a + k_0 n_5 b)}, \quad (S2.5)$$

where it is assumed that the elementary cell of the photonic crystal consists of two layers with refractive indices $n_4$ and $n_5$ with thicknesses $a$ and $b$, respectively. $K$ is the Bloch wavenumber for the Bloch wave propagating along the positive direction of the z-axis. If the optical absorption is neglected $K$ can be found from the following relation:

$$\exp(iK(a+b)) = \alpha + i(\text{sgn}(\sin(k_0 n_4 a + k_0 n_5 b)))\sqrt{1-\alpha^2}, \tag{S2.6}$$

where

$$\alpha = \cos(k_0 n_4 a)\cos(k_0 n_5 b) - \frac{1}{2}\left(\frac{n_4}{n_5} + \frac{n_5}{n_4}\right)\sin(k_0 n_4 a)\sin(k_0 n_5 b), \tag{S2.7}$$

Eq. (S2.6) is valid for the case $|\alpha| \leq 1$, which is fulfilled for the wavelengths outside the photonic bandgaps. However, the most important case is the bandgaps so let's explore it in details. At this $|\alpha| > 1$, and the Bloch wavenumber can be found from

$$\exp(iK(a+b)) = \alpha - (\text{sgn}\,\alpha)\sqrt{\alpha^2 - 1}. \tag{S2.8}$$

One can see that $\exp(iK(a+b))$ becomes real. It follows from Eq. (S2.5) that $n_3$ is imaginary, and Eq. (S2.1) takes the form:

$$E(z) = A\sqrt{n_2^2 - n_3^2}\cos(n_2 k_0(z-d) + \varphi), \tag{S2.9}$$

where

$$\varphi = -\text{atan}\left(\frac{in_3}{n_2}\right). \tag{S2.10}$$

Therefore,

$$|E(z)|^2 = |A|^2(n_2^2 - n_3^2)\cos^2(n_2 k_0(z-d) + \varphi), \tag{S2.11}$$

Eqs. (S2.10) and (S2.11) fully confirm Eqs. (3) and (4).

The phase $\varphi$ calculated by Eqs. (S2.5), (S2.7), (S2.8) and (S2.10) is shown in Fig. 3c (solid lines). For the pristine film Eqs. (S2.1), (S2.2) and (S2.3) were used. The disagreement with rigorous simulations shown by dots is caused by the fact that rigorous simulations were performed for the inclined incidence similar to the experiments. Also optical absorption was taken into account.

### S3. The deduction of formula for excitation amplitudes $A_n$.

The previous formula for the excitation amplitudes $A_n = \frac{\beta_n \gamma^2 \Delta t}{\omega_n d}\sum_s(H + AMk_s^2)\int_0^d h_s(z)\cdot\theta_n(z)\,dz$, mentioned in Ref. [2], was based on the orthogonality assumption:

$$\int_0^d \theta_m(z)\theta_n(z)\,dz \approx \frac{d}{\beta_n}\delta_{mn} \quad (\beta_{n=0} = 1, \quad \beta_{n\neq 0} = 2). \tag{S3.1}$$

However, this is valid only for not large values of the pinning parameter ($\xi d < 1.5 - 2.0$). We would like to have the more complete theory, which will be valid for any values of the pinning parameter. In this case, we should replace Eq. (S3.1) by the following:

$$\int_0^d \varphi_m(z)\theta_n(z)\,dz = \frac{d}{\beta_n'}\delta_{mn}. \tag{S3.2}$$

Here the constants $\beta'_n$ may be found precisely, as will be shown later. Eq. (S3.2) represents an analogue of orthogonality condition, but one more basis $\varphi_n$ (biorthogonal basis) is used to satisfy the condition, because $\theta_n$ functions are not orthogonal to each other in a general case. We will prove the validity of the Eq. (S3.2) below.

Let us write the linearized LLG-equations (Eq. (S3.2)) for the separate modes $\theta_m(z,t) = \theta_m(t) \cdot \theta_m(z)$ and $\varphi_m(z,t) = \varphi_m(t) \cdot \varphi_m(z)$. Neglecting the damping and considering that $\theta_m(t) \sim \sin \omega_m t$, $\varphi_m(t) \sim \cos \omega_m t$, we will obtain:

$$\begin{cases} \omega_m \theta_m(z) = \gamma H \varphi_m(z) - \gamma AM \varphi''_m(z), \\ \omega_m \varphi_m(z) = \gamma \widetilde{H} \theta_m(z) - \gamma AM \theta''_m(z). \end{cases} \quad (S3.3)$$

We will multiply the first equation in (S3.3) by $\varphi_n(z)$ and the second – by $\theta_n(z)$. Then, integrating them and introducing the notations like $\int_0^d \varphi_m(z)\theta_n(z)\,dz = (\varphi_m\,\theta_n)$, we will have:

$$\begin{cases} \omega_m(\theta_m\,\varphi_n) = \gamma H(\varphi_m\,\varphi_n) - \gamma AM(\varphi''_m\,\varphi_n), \\ \omega_m(\varphi_m\,\theta_n) = \gamma \widetilde{H}(\theta_m\,\theta_n) - \gamma AM(\theta''_m\,\theta_n). \end{cases} \quad (S3.4)$$

We can transform the last terms in equations (S3.4) with the use of double integration by parts:

$$\begin{cases} (\varphi''_m\,\varphi_n) = (\varphi'_m\varphi_n - \varphi_m\varphi'_n)\Big|_0^d + (\varphi''_n\,\varphi_m), \\ (\theta''_m\,\theta_n) = (\theta'_m\theta_n - \theta_m\theta'_n)\Big|_0^d + (\theta''_n\,\theta_m). \end{cases} \quad (S3.5)$$

Using the boundary condition (S1.4), we will get from Eq. (S3.5): $(\varphi''_m\,\varphi_n) = (\varphi''_n\,\varphi_m)$, $(\theta''_m\,\theta_n) = (\theta''_n\,\theta_m)$. Taking into account that $(\varphi_m\,\varphi_n) = (\varphi_n\,\varphi_m)$, $(\theta_m\,\theta_n) = (\theta_n\,\theta_m)$, we see that the right part of the equations (S3.4) is invariant under the permutation of indices $m,n$, which, thus, also leads to the invariance of the left part of the equations (S3.4):

$$\begin{cases} \omega_m(\theta_m\,\varphi_n) = \omega_n(\theta_n\,\varphi_m), \\ \omega_m(\varphi_m\,\theta_n) = \omega_n(\varphi_n\,\theta_m). \end{cases} \quad (S3.6)$$

$$\begin{cases} (\theta_m\,\varphi_n) = \dfrac{\omega_n}{\omega_m}(\theta_n\,\varphi_m), \\ (\theta_n\,\varphi_m) = \dfrac{\omega_n}{\omega_m}(\theta_m\,\varphi_n). \end{cases} \quad (S3.7)$$

Substituting the second equation into the first in (S3.7), we will finally have:

$$(\theta_m\,\varphi_n) = \dfrac{\omega_n^2}{\omega_m^2}(\theta_m\,\varphi_n). \quad (S3.8)$$

From Eq. (S3.8) it follows that in a case when $\dfrac{\omega_n}{\omega_m} \neq 1$ (which is valid, when $m \neq n$), the term $(\theta_m\,\varphi_n)$ is equal to zero. This means, that functions $\varphi_n$ constitute a biorthogonal basis to functions $\theta_n$: $(\theta_m\,\varphi_n) \sim \delta_{mn}$, which proves the validity of the Eq. (S3.2). Using this fact, we can deduce the more precise formula for the excitation amplitudes of different PSSW modes.

For obtaining the expression for the excitation amplitudes we can use the initial condition on $\dot{\theta}$:

$$\dot{\theta}(z, t = \Delta t) = \sum_{n=0}^{\infty} A_n \omega_n \theta_n(z) = \gamma^2 \Delta t[H \cdot h(z) - AM \cdot h''(z)]. \quad (S3.9)$$

In a real experiment the IFE-field may be proportional to the superposition of the harmonic functions $h_s(z) \sim \sin k_s z$, $\cos k_s z$ (including $k_s = 0$): $h(z) = \sum_s h_s(z)$. In this case, the Eq. (S3.9) may be simplified using $h_s''(z) = -k_s^2 \cdot h_s(z)$:

$$\sum_{n=0}^{\infty} A_n \omega_n \theta_n(z) = \gamma^2 \Delta t \sum_{s} [H + AMk_s^2] \cdot h_s(z). \quad (S3.10)$$

Now, we shall use the previously obtained "biorthogonal" condition (Eq. S3.2). Multiplying Eq. (S3.10) by $\varphi_m(z)$ and integrating it by $z$ with the further change of index $m$ by $n$, we will obtain the expression for the amplitudes of modes $A_n$:

$$A_n = \frac{\beta_n' \gamma^2 \Delta t}{\omega_n d} \sum_s (H + AMk_s^2) \int_0^d h_s(z) \cdot \varphi_n(z) \, dz. \quad (S3.11)$$

Here $\beta_n' = \frac{d}{(\theta_n \, \varphi_n)}$. If we express the frequency $\omega_n$ through the wavevector $k$, we will get exactly the form of the equations presented in Ref. [2], while the only difference will be in changing the function $\theta_n(z)$ in the expression for the amplitudes by the function $\varphi_n(z)$.

We shall note, that in a limiting case $\xi d \to 0$, the functions $\varphi_n$ and $\theta_n$ take the same form up to the normalizing coefficient $\sqrt{-b_n}$. That is the reason, why the previous approach for calculating the excitation amplitudes was still applicable for the small values of $\xi d$ product.

However, there is one more, even more simple way to obtain the expression for the excitation amplitudes of PSSWs. Let us now use the initial condition (Eq. (S1.3)) on $\varphi$:

$$\varphi(z, t = \Delta t) = \sum_{n=0}^{\infty} A_n \varphi_n = \gamma \Delta t \cdot h(z). \quad (S3.12)$$

Here we used the fact, that the modes $\varphi_n$ and $\theta_n$ are not independent and, thus, have the same amplitudes $A_n$. In order to find these amplitudes, we should multiply Eq. (S3.12) by $\theta_m(z)$ (but not by $\varphi_m(z)$, as was shown before) and integrate it by $z$ with the further change of index $m$ by $n$. With the help of the previously obtained "biorthogonal" condition (Eq. (S3.2)) we will get the expression for the amplitudes of modes $A_n$ (which coincides with the aforementioned Eq.(S1.8)):

$$A_n = \frac{\gamma \Delta t}{(\varphi_n \, \theta_n)} \int_0^d h(z) \cdot \theta_n(z) \, dz = \gamma \Delta t \frac{(\theta_n \, h)}{(\theta_n \, \varphi_n)}. \quad (S3.13)$$

Here $(\varphi_n \, \theta_n) = \int_0^d \varphi_n(z) \cdot \theta_n(z) \, dz$ and can be estimated for small values of $\xi d$ as: $(\varphi_n \, \theta_n) = \frac{\sqrt{-b_n}}{\beta_n} d = \frac{d}{\beta_n} \sqrt{\frac{\widetilde{H} + AMk_n^2}{H + AMk_n^2}}$, $(\beta_{n=0} \approx 1, \beta_{n \neq 0} \approx 2, k_{n=0}^2 = -\chi_{n=0}^2)$.

**References:**

1. A. G. Gurevich and G. A. Melkov, *Magnetization Oscillations and Waves* (CRC Press, 2020).

2. V. A. Ozerov, D. A. Sylgacheva, M. A. Kozhaev, T. Mikhailova, V. N. Berzhansky, M. Hamidi, A. K. Zvezdin, and V. I. Belotelov, "One-dimensional optomagnonic microcavities for selective excitation of perpendicular standing spin waves," J Magn Magn Mater **543**, 168167 (2022).